\RequirePackage{lineno}
\documentclass[aps,prc,twocolumn, floatfix]{revtex4}
\usepackage{graphicx}
\usepackage{dcolumn}
\usepackage{bm}
\usepackage{multirow}
\usepackage[colorlinks=true, linkcolor=blue, citecolor=blue, urlcolor=blue]{hyperref}

\begin{document}


\title{Probing $\Lambda$ potential via its $v_{2}$ flow in hypernuclei-induced reaction}
\author{Gao-Chan Yong$^{1,2,3}$}

\affiliation{
$^1$Institute of Modern Physics, Chinese Academy of Sciences, Lanzhou 730000, China\\
$^2$School of Nuclear Science and Technology, University of Chinese Academy of Sciences, Beijing 100049, China\\
$^3$State Key Laboratory of Heavy Ion Science and Technology, Institute of Modern Physics, Chinese Academy of Sciences, Lanzhou 730000, China
}

\begin{abstract}

The hyperon potential, particularly its behavior at high densities, is crucial for resolving the ``hyperon puzzle'' in neutron stars and for advancing our understanding of the strong interactions between strange and non-strange particles in high baryon density environments. Using the hadronic transport model AMPT-HC, hypernucleus-nucleus collision is studied. It is found that at beam energies below the threshold for hyperon production, the hyperon elliptic flow exhibits noticeable asymmetry between the positive and negative rapidity regions and is sensitive to the strength of the hyperon potential, especially in the large negative rapidity region. One can extract the hyperon potential approximately twice the saturation density based on the hyperon elliptic flow in the negative rapidity region, and the hyperon potential around the saturation density based on the hyperon elliptic flow in the positive rapidity region.

\end{abstract}

\maketitle

%
Neutron stars, traditionally thought to consist mainly of neutrons \cite{prak1997}, may also contain strange matter, challenging conventional views and reshaping our understanding of their nature \cite{latt2004,Weber2005}. This hypothesis has profound implications for their internal structure, thermal evolution, and observable behavior \cite{nkg2001}. For instance, strange particles could soften the equation of state (EoS) of dense matter \cite{universe2021}, altering core-collapse supernova mechanisms \cite{npa1997,soft2,soft3,soft4}, influencing cooling processes \cite{cool92,cool99,cool3,cool4,cool5}, and affecting merger dynamics, potentially generating unique gravitational wave signals detectable by advanced observatories \cite{mark21,gw17}. The study of strangeness in neutron stars has thus drawn interest from astrophysics, particle physics, and nuclear physics \cite{apj85,sch96,nk82,lon15,ch16,wz12,ger20,jj23}, aiming to uncover insights into the strong nuclear force and extreme matter conditions unattainable in terrestrial labs \cite{wz12,jj23,sxx2023,vid18,ppnp2020,chenjh}.

The so-called ``hyperon puzzle'' in neutron stars \cite{puzzle16,universe2021,vid18} arises from the contradiction between theoretical models, which predict that hyperons would soften the equation of state (EoS), leading to the formation of lower mass neutron stars, and the recent observations of neutron stars with significantly higher masses \cite{lon15}. To address the ``hyperon puzzle'', researchers employ diverse theoretical frameworks, including nuclear many-body theories \cite{ch16,ppnp2020,mdi1998,mas2015}, effective field theories \cite{hai2020,jm14,geng2022}, and perturbative quantum chromodynamics \cite{ms1987,xiacj2017}, alongside astrophysical observations \cite{longwh12,hell14}. These approaches collectively advance understanding of strange matter and the strong nuclear force in dense environments.

Despite the availability of various theoretical methodologies, the interactions between strange and non-strange particles in nuclear matter remain shrouded in significant theoretical uncertainties. These interactions have rarely been directly probed in terrestrial nuclear experiments. As highlighted in Ref.~\cite{yongprd2023}, a promising approach to resolving the ``hyperon puzzle'' lies in the integration of transport model simulations with experimental data from global nuclear facilities. This suggests that nuclear reactions could serve as a powerful tool for investigating hyperon-nucleon interactions within the nuclear medium. The production of hypernuclei can serve as a valuable tool for assessing the $\Lambda$ potential in the medium near saturation density \cite{incl2018}. Similarly, analyzing the directed flows of hyperons or hypernuclei in typical heavy-ion collisions offers critical insights into the $\Lambda$ potential in the medium at high densities \cite{star2023}. However, such studies are accompanied by significant theoretical uncertainties, primarily stemming from the production mechanisms of $\Lambda$ particles, the formation of hypernuclei, and the influence of non-strange baryon potentials, given that $\Lambda$ hyperons are secondary particles in heavy-ion collisions. Recently, it has been observed that, experiments involving $\Lambda$ interactions can be conducted using secondary-beam facilities in terrestrial laboratories, such as those at JLab/CLAS \cite{clas2019} and BEPCII/BESIII \cite{bas31,bas32}. Additionally, reactions involving $\Lambda$-hypernuclei could potentially be carried out at GSI/FAIR, utilizing a specialized HYDRA (HYpernuclei Decay at R3B Apparatus) time-projection chamber prototype \cite{gsi23}. Experimental plans for such research are currently being developed. Therefore, hyperon-nucleus scattering experiments provide insights into these interactions at densities close to saturation, while hypernucleus-nucleus collisions extend the exploration to higher densities \cite{yongplb2024}. By circumventing the complexities associated with the generation of $\Lambda$-hyperons or $\Lambda$-hypernuclei in the medium, such research approaches significantly reduce theoretical uncertainties.

In this work, I focused on semi-central hypernucleus-nucleus collisions at a beam energy of 400 MeV in the laboratory frame to analyze the elliptic flow ($v_{2}$) of pre-existing, non-secondarily produced $\Lambda$ hyperons. The results reveal that the rapidity and transverse momentum dependence of the $\Lambda$ elliptic flow are highly sensitive to the strength of the hyperon potential. Notably, the $\Lambda$ elliptic flow in the negative rapidity region reflects the potential strength at high densities, whereas the flow in the positive rapidity region probes the potential near saturation density at lower densities. This dual sensitivity demonstrates that the $\Lambda$ elliptic flow in semi-central hypernucleus-nucleus collisions can effectively map the strength of the $\Lambda$ potential across a wide range of densities, from high to low.

%
Consistent with the objectives of this research, the multi-phase transport (AMPT) model \cite{AMPT2005} has been recently enhanced to include not only simulations that account for both partonic and hadronic degrees of freedom but also to conduct pure hadron cascade simulations incorporating hadronic mean-field potentials (AMPT-HC) \cite{nst2021,cas2021,yongrcas2022,yongplb2023,phase2024}. The latest iteration of the AMPT-HC model integrates the initial density and momentum distributions of nucleons within colliding nuclei, as determined by Skyrme-Hartree-Fock calculations utilizing Skyrme M$^{\ast}$ force parameters \cite{skyrme86}, and the Fermi momentum derived from the local density via the local Thomas-Fermi approximation. This model also reflects recent experimental insights into the nucleon momentum distribution, featuring a high-momentum tail that reaches up to approximately twice the local Fermi momentum \cite{yongsrc}. The potentials for nucleons, resonances, hyperons, and their antiparticles are implemented using the test-particle method \cite{yongrcas2022}. The density-dependent single nucleon mean-field potential currently in use aligns with that described in Ref.~\cite{cas2021}. The kaon potential is sourced from Ref.~\cite{ligq97}, while the pion potential is considered negligible at relatively high energies \cite{pionp15}. For strange baryons, we apply the quark counting rule, which suggests that these baryons interact with other baryons exclusively through their non-strange components \cite{mos74,chung2001}. Consequently, the hyperon potential currently employed is solely of a density-dependent form --- specifically, a Skyrme-type potential that is two-thirds the strength of the density-dependent single-nucleon potential (the value of the hyperon potential at saturation density is approximately -35 MeV \cite{incl2018}). However, this does not impede the quest for observables that are sensitive to the hyperon potential. The free elastic proton-proton cross section, denoted as $\sigma_{pp}$, and the neutron-proton cross section, denoted as $\sigma_{np}$, are determined by experimental data. The free elastic neutron-neutron cross section, denoted as $\sigma_{nn}$, is assumed to be equivalent to $\sigma_{pp}$ at a similar center of mass energy. Additionally, it is presumed that all other baryon-baryon free elastic cross sections are equivalent to the nucleon-nucleon elastic cross section at the same center of mass energy. An experimental energy-dependent nucleon-nucleon inelastic total cross-section is utilized at lower energies \cite{nninel87}. It is noted here that subsequent studies, as shown in Figure 4, indicate that the values of the nucleon-nucleon and hyperon-nucleon scattering cross-sections do not have significant impacts on the hyperon elliptic flow in hypernucleus-nucleus collisions. Consequently, I made the assumption of a free nucleon-nucleon scattering cross-section, as well as the supposition that the hyperon-nucleon scattering cross-section is equivalent to that of the nucleon-nucleon scattering.

%
To uncover the underlying dynamical information from heavy-ion collisions, the differential distribution of particles is typically described using a Fourier series \cite{flow6,flow7,flow8}. The elliptic flow, denoted by $v_{2}$, can be expressed as:
$$
   v_{2} = \langle \cos(2\phi) \rangle = \left\langle \frac{p_{x}^{2} - p_{y}^{2}}{p_{T}^{2}} \right\rangle,
$$
where $p_{T} = \sqrt{p_{x}^{2} + p_{y}^{2}}$ represents the transverse momentum of the particle, and $\phi$ corresponds to its azimuthal angle. In the following section, I will focus on exploring the elliptic flow $v_{2}$ of $\Lambda$ particles and its associated aspects.

\begin{figure}[t]
\centering
\includegraphics[width=0.45\textwidth]{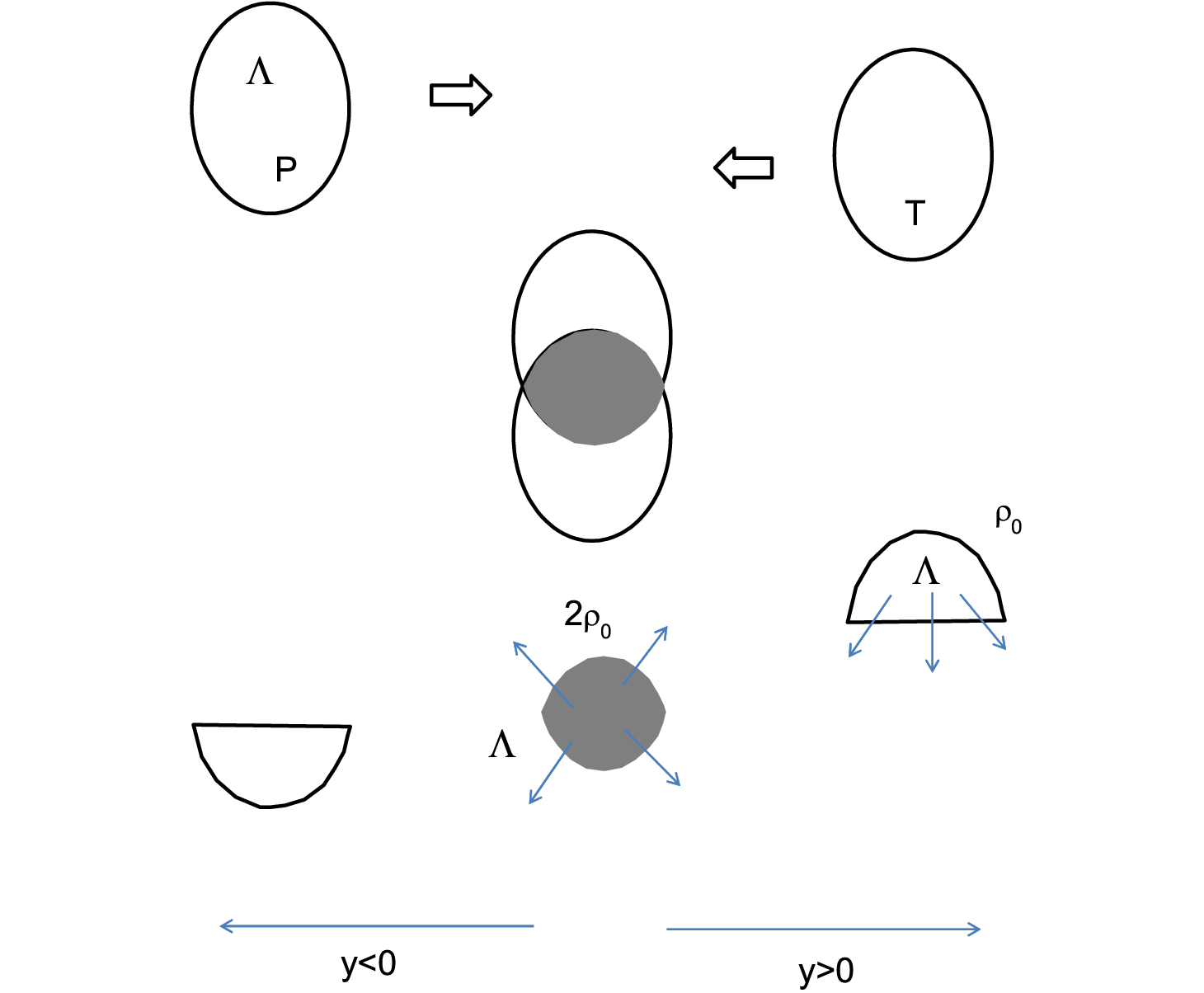}
\vspace{0.25cm}
\caption{Schematic diagram of the collision reaction between a $\Lambda$-hypernucleus and a normal nucleus with the beam energy employed in the text.} \label{sketch}
\end{figure}
To intuitively understand the origin of hyperons in hypernucleus-nucleus collisions, I present a schematic diagram of $\Lambda$-hypernucleus-nucleus collisions as demonstrated in Figure~\ref{sketch}. It demonstrates that when a $\Lambda$-hypernucleus, acting as a projectile, collides with a normal atomic nucleus, the spectator fragments of the projectile primarily contribute to the $\Lambda$ hyperon distribution at positive rapidities, thus mainly reflecting the $\Lambda$ potential effects in the low-density saturation region. Whereas the participant part of the projectile nucleus contributes to the $\Lambda$ hyperon distribution at both negative and positive rapidities, mainly reflecting the $\Lambda$ potential effects in the region of twice the saturation nuclear density with the beam energy studied here. In the following research, I will not employ central collisions between hypernuclei and normal nuclei, using the transverse momentum distribution or kinetic energy distribution of hyperons to probe the strength of the hyperon potential. This is primarily because the transverse momentum distribution or kinetic energy distribution of hyperons, compared to collective flow observables, exhibits more theoretical uncertainties. Additionally, I will not use the hyperon directed flow $v_1$ to investigate the hyperon potential strength in semi-central collisions between hypernuclei and normal nuclei. This is because my research has found that the strength of the hyperon directed flow $v_1$, such as its slope near mid-rapidity, is not very sensitive to the hyperon potential strength somehow. Therefore, in the following research, I will focus exclusively on the study of hyperon elliptic flow $v_2$.

For the selection of hypernuclei, to facilitate the study, I have chosen to use heavy hypernuclei colliding with heavy atomic nuclei. This approach ensures the formation of a large volume of dense nuclear matter including hyperon. Of course, other heavy hypernuclei colliding with arbitrary heavy ions can also be considered. For nuclear reactions induced by very light hypernuclei, the small volume of the produced hyperon nuclear matter may result in a short duration of the hyperon potential interaction, thereby making the hyperon potential effect less pronounced. Regarding the choice of collision energy, I recommend keeping the beam energy below the threshold energy for $\Lambda$ hyperon production. For instance, in this study, a laboratory beam energy of 400 MeV is used, ensuring that all hyperons in the final state originate from the original hypernuclei. This approach avoids the uncertainties associated with inelastic channels for hyperon production.

\begin{figure}[t]
\centering
\vspace{-0.25cm}
\includegraphics[width=0.45\textwidth]{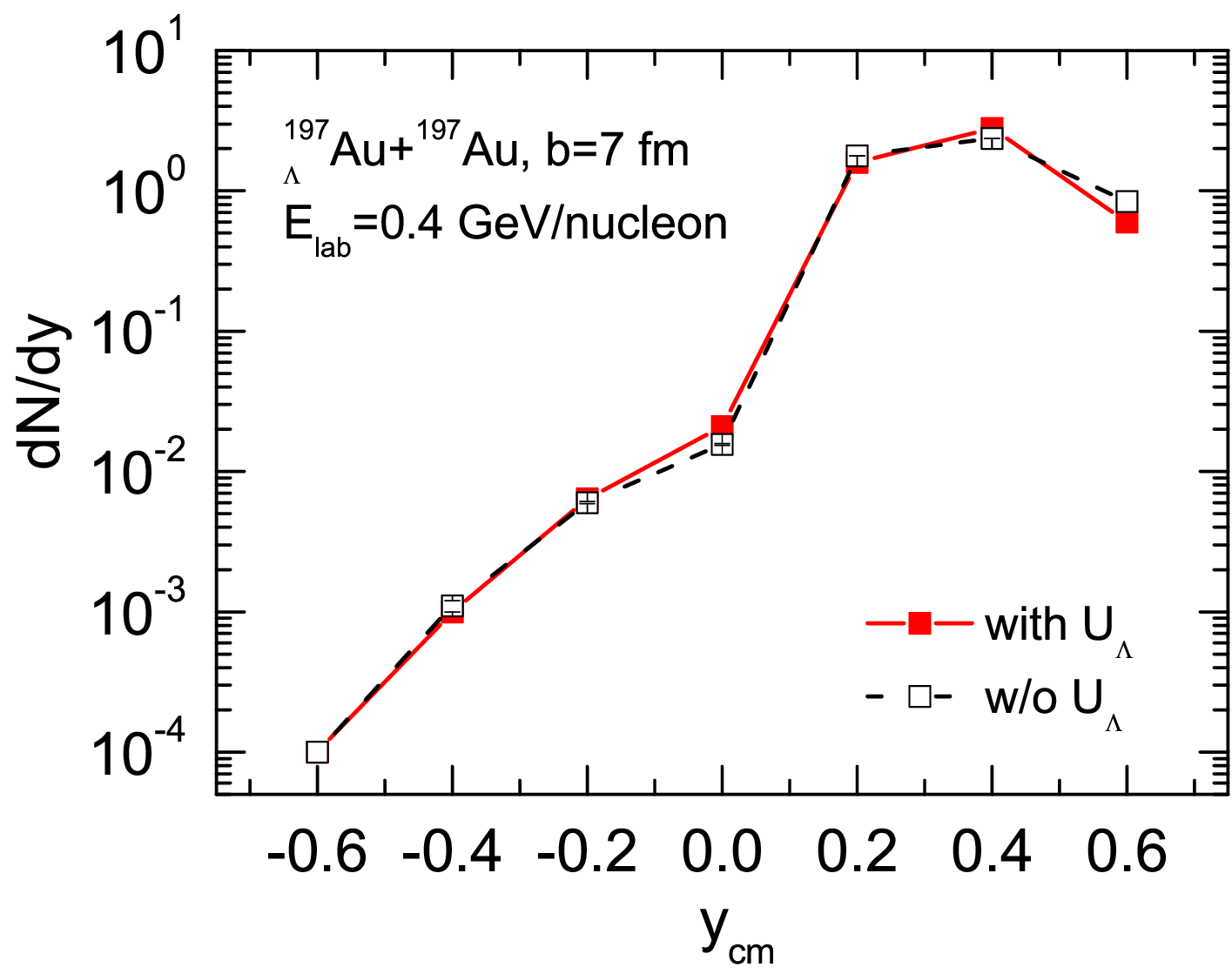}
\caption{Rapidity distribution of original $\Lambda$'s in the center-of-mass system and the effects of $\Lambda$ potential in semi-central collisions of $^{197}_{\Lambda}$Au + Au at a laboratory beam energy of 400 MeV.} \label{dndy}
\end{figure}
Before presenting the elliptic flow of $\Lambda$ hyperons, I will first display the rapidity distribution of $\Lambda$ hyperons in $\Lambda$-hypernucleus-nucleus collisions at a laboratory beam energy of 400 MeV. Figure~\ref{dndy} illustrates the rapidity distribution of $\Lambda$ hyperons in semi-central collisions between hypernuclei and atomic nuclei. It is evident that the majority of $\Lambda$ hyperons are distributed in the positive rapidity region, while $\Lambda$ hyperons are scarce in the negative rapidity region. This is primarily because, in the case of large impact parameters, most $\Lambda$ hyperons, or rather, $\Lambda$ hyperons are more likely to remain in the ``spectator'' fragments of the projectile, with only a small fraction of $\Lambda$ hyperons entering the ``participant'' matter. Since there are no $\Lambda$ hyperons in the target ``spectator'' fragments, this results in a significantly lower distribution of $\Lambda$ hyperon in the negative rapidity region. The dynamic or kinematic behavior of $\Lambda$ hyperons from different sources actually reflects the $\Lambda$ potential in different density regions. However, it is evident that the rapidity distribution of the $\Lambda$ hyperon remains largely unaffected by variations in the strength of the $\Lambda$ potential. Upon closer examination, it becomes apparent that in the vicinity of the center-of-mass rapidity y$_{cm}$ = 0.4 and around y$_{cm}$ = 0, the number of $\Lambda$ hyperons in the presence of a $\Lambda$ potential exceeds that in the absence of such a potential. This phenomenon is likely attributed to the compression of hyperonic matter during nucleus-nucleus collisions, coupled with the repulsive nature of the $\Lambda$ hyperon potential.

\begin{figure}[t!]
\centering
\vspace{-0.0cm}
\includegraphics[width=0.45\textwidth]{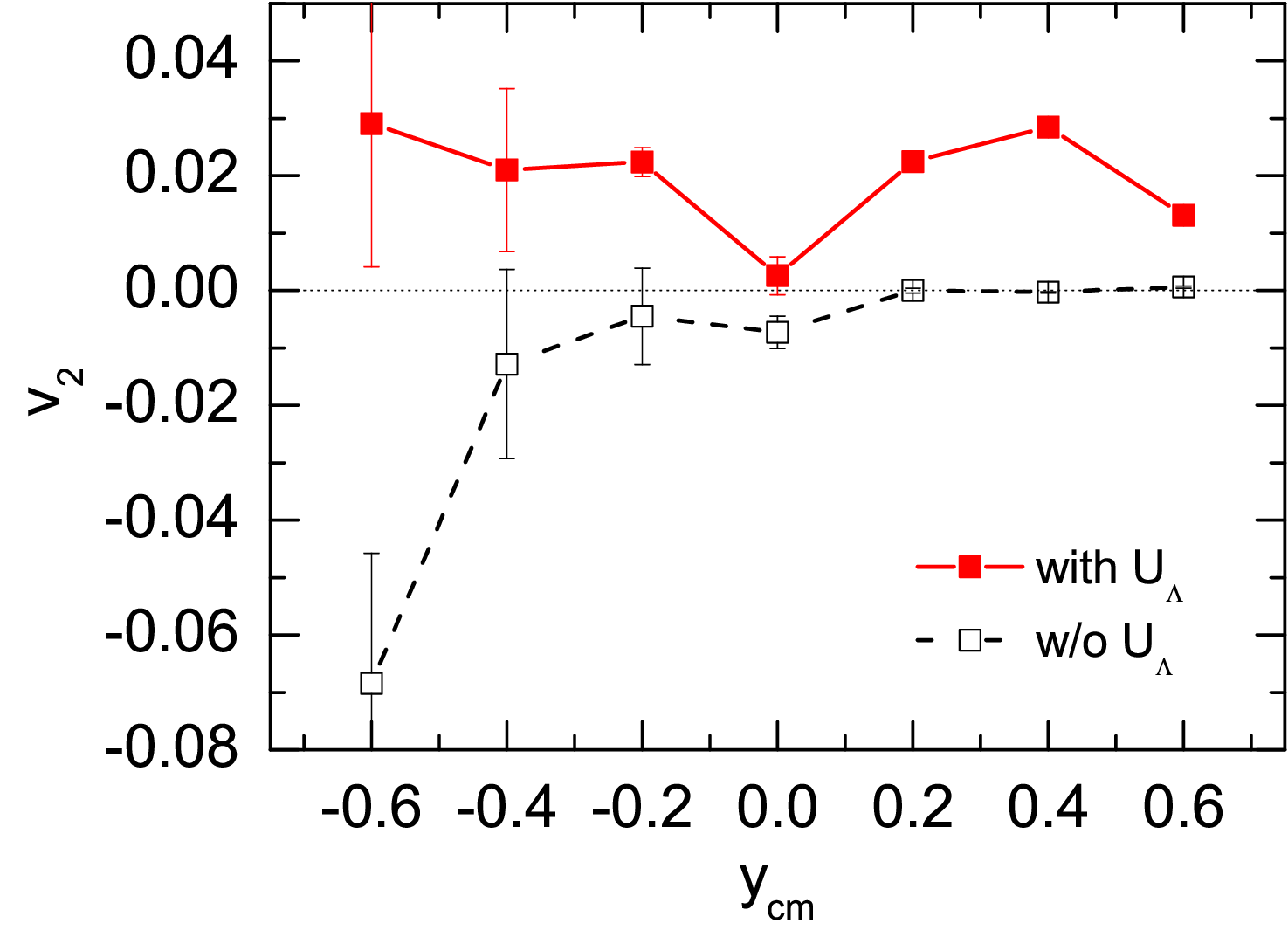}
\caption{Variation of $\Lambda$ elliptic flow $v_{2}$ with rapidity and the effects of $\Lambda$ potential in the same semi-central $^{197}_{\Lambda}$Au + Au collisions at E$_{lab}$= 400 MeV.} \label{v2y}
\end{figure}
I will now shift my focus to the study of $\Lambda$ elliptic flow $v_{2}$. Figure~\ref{v2y} illustrates the variation of $\Lambda$ elliptic flow $v_{2}$ with rapidity in hypernucleus-nucleus collisions. It is evident that the hyperon elliptic flow $v_{2}$ is highly sensitive to the hyperon potential, both in the positive and negative rapidity regions, particularly in the large negative rapidity region. In the positive rapidity region, since hyperons primarily originate from the ``spectator'' fragments of the projectile, the hyperon potential strength around the saturation nuclear density is predominantly probed. In the presence of a hyperon potential, the hyperon elliptic flow $v_{2}$ assumes a positive value and is significantly larger than in the absence of such a potential. Without the hyperon potential, the hyperon elliptic flow $v_{2}$ essentially vanishes to zero. This is likely due to the compression of the ``spectator'' hyperonic matter in the $x$ direction, which enhances the repulsion of the hyperon potential in the $x$ direction. Turning to the negative rapidity region, it is observed that the hyperon elliptic flow $v_{2}$ becomes even more sensitive to the hyperon potential strength, especially in the large negative rapidity region. This is attributed to the greater compression of the ``participant'' hyperonic matter, leading to a stronger repulsion from the hyperon potential. In the absence of a hyperon potential, the negative elliptic flow $v_{2}$ indicates that hyperonic matter is primarily emitted in the $y$ direction. However, with the hyperon potential present, the elliptic flow $v_{2}$ becomes positive again. This reaffirms that, at the current beam energy, the hyperon potential primarily repels hyperons in the $x$ direction. This is likely because the compressed, high-density ``participant'' hyperonic matter experiences greater compression in the $x$ direction, resulting in a larger density gradient in that direction. Given that the number of hyperons in the positive rapidity region is significantly higher than in the negative rapidity region, the hyperon elliptic flow $v_{2}$ in the negative rapidity region exhibits large statistical uncertainties. Since the hyperon elliptic flow in the positive and negative rapidity regions probes the hyperon potential strength at different density regimes, the elliptic flow in these regions holds distinct research significance.

\begin{figure}[t]
\centering
\vspace{-0.1cm}
\includegraphics[width=0.45\textwidth]{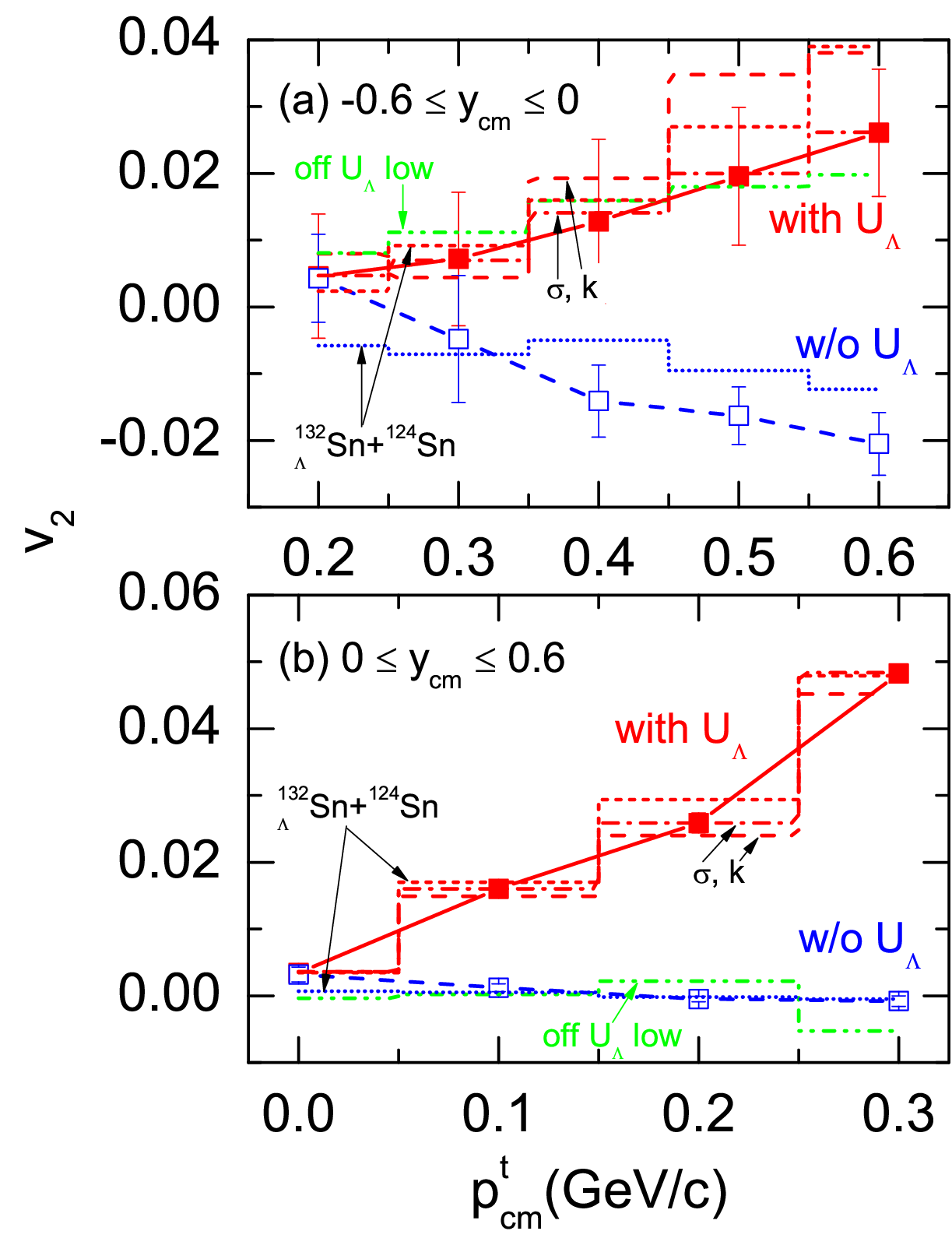}
\caption{Variation of $\Lambda$ elliptic flow $v_{2}$ strength with the transverse momentum $p^{t}_{cm}$ distribution in the center-of-mass system and the effects of $\Lambda$ potential in the same semi-central $^{197}_{\Lambda}$Au + Au collisions at E$_{lab}$= 400 MeV. The upper panel (a) shows the situation in the negative rapidity region, while the lower panel (b) displays the situation in the positive rapidity region. The meanings of the various step lines in the figure are as follows: $\sigma$ indicates that the baryon-baryon elastic scattering cross-section is reduced by half; $k$ signifies that the incompressibility coefficient is doubled; $off U_{\Lambda} low$ denotes the removal of the hyperon potential below the saturation nuclear density; and $Sn+Sn$ represents the alteration of the reaction system for hypernucleus-nucleus collisions.} \label{v2pt}
\end{figure}
To explore the hyperon elliptic flow in greater detail and comprehensiveness, I further present the transverse momentum distribution of the hyperon elliptic flow. Figure~\ref{v2pt} illustrates the distribution of $\Lambda$ elliptic flow as a function of transverse momentum in both positive and negative rapidity regions. It is evident that the transverse momentum distribution of hyperon elliptic flow is highly sensitive to the $\Lambda$ potential, regardless of whether it is in the negative or positive rapidity region. As previously discussed, although the number of $\Lambda$ hyperons in the large negative rapidity region is small, this region holds significant research value for probing the high-density hyperon potential. According to the RMF theory \cite{rmf2015}, since the $\rho$ meson-nucleon coupling constant $g_{\rho\Lambda}$ in the vector component of the Lambda hyperon potential is approximately zero, the isospin-density-dependent part of the Lambda vector potential thus contributes negligibly to the overall Lambda hyperon potential. Therefore, the high-density hyperon potential extracted from heavy-ion collision experiments can be applied to neutron star environments with high density and isospin density ($\rho_{n}-\rho_{p}$).
The investigation into the $\Lambda$ hyperon potential in dense nuclear environments is critically important for understanding neutron stars, especially when it comes to solving the ``hyperon puzzle''. Therefore, conducting hypernucleus-nucleus collision experiments in ground-based laboratories to probe and constrain the high-density hyperon potential is of paramount importance for addressing the ``hyperon puzzle'' in neutron stars and for understanding the strong interactions between strange and non-strange particles in high baryon density environments.

As previously discussed, the hyperon potential can be extracted by studying the production of secondary particle hyperons, hyperon collective flow, and hypernuclei production based on typical nucleus-nucleus collisions \cite{star2023}. However, such studies are subject to significant uncertainties. On one hand, the production of hyperons, hyperon collective flow, and hypernuclei production are all influenced by the inelastic cross-sections of hyperon production and the theoretical uncertainties in the coalescence of hyperons with other nucleons to form hypernuclei. On the other hand, since hyperons are secondary particles, their production and collective flow are inevitably affected by the stiffness of the nuclear matter EoS. The stiffness of the high-density nuclear matter EoS is one of the forefront topics in current nuclear physics research, and thus its stiffness is highly uncertain. As can be seen from Figure~\ref{v2pt}, the primordial hyperon's elliptic flow observable proposed in this study is insensitive to the baryon-baryon scattering cross-section, the stiffness of the EoS, and the reaction system, but is sensitive only to the strength of the hyperon potential. Additionally, Figure~\ref{v2pt} also shows that the negative rapidity hyperon elliptic flow indeed probes the high-density behavior of the hyperon potential, while the positive rapidity hyperon elliptic flow probes the low-density behavior of the hyperon potential.

%
To summarize, based on the hadronic transport model AMPT-HC, the semi-central $^{197}_{\Lambda}$Au + Au collision at a beam energy of 400 MeV per nucleon in the laboratory frame is studied. It is found that the number of $\Lambda$ hyperons in the positive rapidity region of the center-of-mass system is significantly higher than that in the negative rapidity region. The $\Lambda$ hyperon elliptic flow $v_{2}$, whether in terms of rapidity or transverse momentum distributions, is sensitive to the $v_{2}$ hyperon potential in both positive and negative rapidity regions, particularly in the large negative rapidity region. Therefore, the $\Lambda$ potential around twice the saturation density can be extracted based on the $\Lambda$ elliptic flow in the negative rapidity region, while the $\Lambda$ potential near approximately the saturation density can be extracted based on the $\Lambda$ elliptic flow $v_{2}$ in the positive rapidity region. The study of the density dependence of the $\Lambda$ potential is crucial for resolving the ``hyperon puzzle'' in neutron stars and for advancing our understanding of the strong interactions between strange and non-strange particles. Therefore, we hope that the experimental research on hypernucleus-nucleus collisions will be implemented as soon as possible at GSI/FAIR and other similar large-scale scientific facilities internationally. Experimental challenges do exist, but only through such efforts can the high-density hyperon potential be cleanly extracted.

%
This work is supported by the National Natural Science Foundation of China under Grant Nos. 12275322, 12335008 and CAS Project for Young Scientists in Basic Research YSBR-088.


\begin{thebibliography}{100}
\bibitem{prak1997}Madappa Prakash, Ignazio Bombaci, Manju Prakash, Paul J. Ellis, James M. Lattimer, Roland Knorren, Phys. Rept. {\bf 280}, 1 (1997).
\bibitem{latt2004}J. M. Lattimer, M. Prakash, Science {\bf 304}, 536 (2004).
\bibitem{Weber2005}F. Weber, Prog. Part. Nucl. Phys. {\bf 54}, 193 (2005).
\bibitem{nkg2001}N. K. Glendenning, Phys. Rep. {\bf 342},  393 (2001).
\bibitem{universe2021}I. Vida\~{n}a, Universe 2021, {\bf 7}, 376 (2021).
\bibitem{npa1997}Shmuel Balberg, Avraham Gal, Nucl. Phys. A {\bf 625}, 435 (1997).
\bibitem{soft2}Adriana R. Raduta, Micaela Oertel, Armen Sedrakian, Mon. Not. Roy. Astron. Soc. {\bf 499}, 914 (2020).
\bibitem{soft4}Adriana R. Raduta, Eur. Phys. J. A {\bf 58}, 115 (2022).
\bibitem{soft3}Armen Sedrakian, Arus Harutyunyan, Eur. Phys. J. A {\bf 58}, 137 (2022).
\bibitem{cool92}M. Prakash, M. Prakash, J. M. Lattimer, and C. J. Pethick, Astrophys. J. {\bf 390}, L77 (1992).
\bibitem{cool99}J. A. Pons, S. Reddy, M. Prakash, J. M. Lattimer, and J. A. Miralles, Astrophys. J. {\bf 513}, 780 (1999).
\bibitem{cool4}Adriana R. Raduta, Armen Sedrakian, Fridolin Weber, Mon. Not. Roy. Astron. Soc. {\bf 475}, 4347 (2018).
\bibitem{cool3}Adriana R. Raduta, Jia Jie Li, Armen Sedrakian, Fridolin Weber, Mon. Not. Roy. Astron. Soc. {\bf 487}, 2639 (2019).
\bibitem{cool5}F. Anzuini, A. Melatos, C. Dehman, D. Vigan\`{o}, J. A. Pons, Mon. Not. Roy. Astron. Soc. {\bf 509}, 2609 (2021).
\bibitem{gw17}B. P. Abbott \emph{et al}. (LIGO Scientific Collaboration and Virgo Collaboration),
Phys. Rev. Lett. {\bf 119}, 161101 (2017).
\bibitem{mark21}Mark G. Alford and Alexander Haber, Phys. Rev. C {\bf 103}, 045810 (2021).
\bibitem{nk82}Norman K. Glendenning, Phys. Lett. B {\bf 114}, 392 (1982).
\bibitem{apj85}N. K. Glendenning, Astrophys. J. {\bf 293}, 470 (1985).
\bibitem{sch96}J. Schaffner, I. N. Mishustin, Phys. Rev. C {\bf 53}, 1416 (1996).
\bibitem{wz12}Wei-Zhou Jiang, Bao-An Li, and Lie-Wen Chen, Astrophys. J. {\bf 756}, 56 (2012).
\bibitem{lon15}Diego Lonardoni, Alessandro Lovato, Stefano Gandolfi, and Francesco Pederiva,
               Phys. Rev. Lett. {\bf 114}, 092301 (2015).
\bibitem{ch16}D. Chatterjee, and I. Vida\~{n}a, Eur. Phys. J. A {\bf 52}, 1 (2016).
\bibitem{ger20}D. Gerstung, N. Kaiser, W. Weise, Eur. Phys. J. A {\bf 56}, 175 (2020).
\bibitem{jj23}A. Sedrakian, J. J. Li, F. Weber, Prog. Part. Nucl. Phys. {\bf 131}, 104041 (2023).
\bibitem{vid18}I. Vida\~{n}a, Proc. R. Soc. Lond. A {\bf 474}, 0145 (2018).
\bibitem{sxx2023}Xiangdong Sun, Zhiqiang Miao, Baoyuan Sun, Ang Li, Astrophys. J. {\bf 942}, 55 (2023).
\bibitem{ppnp2020}L. Tolos, L. Fabbietti, Prog. Part. Nucl. Phys. {\bf 112}, 103770 (2020).
\bibitem{chenjh}J. Chen, D. Keane, Y.G. Ma, A. Tang, Z. Xu, Phys. Rept. {\bf 760}, 1 (2018).
\bibitem{puzzle16}D. Chatterjee, I. Vida\~{n}a, Eur. Phys. J. A {\bf 52}, 29 (2016).
\bibitem{mdi1998}H.-J. Schulze, M. Baldo, U. Lombardo, J. Cugnon, and A. Lejeune, Phys. Rev. C {\bf 57}, 704 (1998).
\bibitem{mas2015}K. A. Maslov, E. E. Kolomeitsev, D.N. Voskresensky, Phys. Lett. B {\bf 748}, 369 (2015).
\bibitem{hai2020}J. Haidenbauer,  U.-G. Mei{\ss}ner, A. Nogga, Eur. Phys. J. A {\bf 56}, 91 (2020).
\bibitem{jm14}J. M. Alarc\'{o}n, L. S. Geng, J. Martin Camalich, J. A. Oller, Phys. Lett. B {\bf 730}, 342 (2014).
\bibitem{geng2022} Jing Song, Zhi-Wei Liu, Kai-Wen Li, and Li-Sheng Geng, Phys. Rev. C {\bf 105}, 035203 (2022).
\bibitem{ms1987}Fridolin Weber, Prog. Part. Nucl. Phys. {\bf 54}, 193 (2005).
\bibitem{xiacj2017}Cheng-Jun Xia, Shan-Gui Zhou, Nucl. Phys. B {\bf 916}, 669 (2017).
\bibitem{longwh12}Wen Hui Long, Bao Yuan Sun, Kouichi Hagino, and Hiroyuki Sagawa, Phys. Rev. C {\bf 85}, 025806 (2012).
\bibitem{hell14}Thomas Hell and Wolfram Weise, Phys. Rev. C {\bf 90}, 045801 (2014).
\bibitem{yongprd2023}Gao-Chan Yong, Phys. Rev. D {\bf 108}, L091507 (2023).
\bibitem{incl2018}J. L. Rodr\'{\i}guez-S\'{a}nchez, J.-C. David, J. Hirtz, J. Cugnon, and S. Leray, Phys. Rev. C {\bf 98}, 021602(R) (2018).
\bibitem{star2023}B. E. Aboona \emph{et al}. (STAR Collaboration), Phys. Rev. Lett. {\bf 130}, 212301 (2023).
\bibitem{clas2019}John W. Price; for the CLAS Collaboration, AIP Conf. Proc. {\bf 2130}, 020004 (2019).
\bibitem{bas31}M. Ablikim \emph{et al}. (BESIII Collaboration), arXiv:2401.09012 (2024).
\bibitem{bas32}Jianping Dai, Hai-Bo Li, Han Miao, Jian-Yu Zhang, arXiv:2209.12601 (2022).
\bibitem{gsi23}S. Velardita, H. Alvarez-Pol, T. Aumann \emph{et al}., Eur. Phys. J. A {\bf 59}, 139 (2023).
\bibitem{yongplb2024}Gao-Chan Yong, Phys. Lett. B {\bf 853}, 138662 (2024).
\bibitem{AMPT2005}Zi-Wei Lin, Che Ming Ko, Bao-An Li, Bin Zhang, Subrata Pal, Phys. Rev. C {\bf 72}, 064901 (2005).
\bibitem{nst2021}Zi-Wei Lin, Liang Zheng, Nucl. Sci. Tech. {\bf 32}, 113 (2021).
\bibitem{cas2021}Gao-Chan Yong, Zhi-Gang Xiao, Yuan Gao, Zi-Wei Lin, Phys. Lett. B {\bf 820}, 136521 (2021).
\bibitem{yongrcas2022}Gao-Chan Yong, Bao-An Li, Zhi-Gang Xiao, and Zi-Wei Lin, Phys. Rev. C {\bf 106}, 024902 (2022).
\bibitem{yongplb2023}Gao-Chan Yong, Phys. Lett. B {\bf 843}, 138051 (2023).
\bibitem{phase2024}Gao-Chan Yong, Phys. Lett. B {\bf 848},138327 (2024).
\bibitem{skyrme86}J. Friedrich and P. G. Reinhard, Phys. Rev. C {\bf 33}, 335 (1986).
\bibitem{yongsrc}Gao-Chan Yong, Phys. Rev. C {\bf 104}, 014613 (2021).
\bibitem{ligq97}G. Q. Li, C.-H. Lee, and G. E. Brown, Phys. Rev. Lett. {\bf 79}, 5214 (1997).
\bibitem{pionp15}W. M. Guo, G. C. Yong, H. Liu, and W. Zuo, Phys. Rev. C {\bf 91}, 054616 (2015).
\bibitem{mos74}S. A. Moszkowski, Phys. Rev. D {\bf 9}, 1613 (1974).
\bibitem{chung2001}P. Chung \emph{et al}. (E895 Collaboration), Phys. Rev. Lett. {\bf 86}, 2533 (2001).
\bibitem{nninel87}J. Bystricky, P. La France, F. Lehar, F. Perrot, T. Siemiarczuk, P. Winternitz, J. Phys. France {\bf 48}, 1901 (1987).
\bibitem{flow6}H. Sorge, Phys. Rev. Lett. {\bf 78}, 2309 (1997).
\bibitem{flow7}H. Sorge, Phys. Lett. B {\bf 402}, 251 (1997).
\bibitem{flow8}J. Y. Ollitrault, Phys. Rev. D {\bf 46}, 229 (1992).
\bibitem{rmf2015}M. Oertel, C. Provid\^{e}ncia, F. Gulminelli and Ad. R. Raduta, J. Phys. G: Nucl. Part. Phys. {\bf 42}, 075202 (2015).

\end{thebibliography}
\end{document}